\documentclass[useAMS]{mn2e}
\usepackage{amssymb}

\usepackage{graphicx}
\title[Eurybates --- the only asteroid family among Trojans?]%
{Eurybates --- the only asteroid family among Trojans?}
\author[M. Bro\v{z} and J. Rozehnal]%
{M. Bro\v{z}$^{1}$\thanks{E-mail: mira@sirrah.troja.mff.cuni.cz (MB)}
and
J. Rozehnal$^{1,2}$
\\
$^{1}$Institute of Astronomy, Charles University, Prague, V Hole\v sovi\v ck\'ach 2, 18000 Prague 8, Czech Republic\\
$^{2}$\v Stef\'anik Observatory, Pet\v r\'in 205, 11800 Prague, Czech Republic
}

\begin{document}

\date{Accepted ???. Received ???; in original form ???}

\pagerange{\pageref{firstpage}--\pageref{lastpage}} \pubyear{2010}

\maketitle

\label{firstpage}

%%%%%%%%%%%%%%%%%%%%%%%%%%%%%%%%%%%%%%%%%%%%%%%%%%%%%%%%%%%%%%%%%%%%%%%%

\begin{abstract}
We study orbital and physical properties of Trojan asteroids of Jupiter.
We try to discern all families previously discussed in literature,
but we conclude there is only one significant family among Trojans,
namely the cluster around asteroid (3548)~Eurybates.
It is the only cluster, which has all of the following characteristics:
(i)~it is clearly concentrated in the proper-element space;
(ii)~size-frequency distribution is different from background asteroids;
(iii)~we have a reasonable collisional/dynamical model of the family.
Henceforth, we can consider it as a real collisional family.

We also report a discovery of a possible family around the asteroid (4709) Ennomos, composed mostly of small asteroids.
The asteroid (4709) Ennomos is known to have a very high albedo $p_V \simeq 0.15$,
which may be related to a hypothetical cratering event which exposed ice
(Fern\'andez et al. 2003). The relation between the collisional family
and the exposed surface of the parent body is a unique opportunity
to study the physics of cratering events. However, more data are needed
to confirm the existence of this family and its relationship with Ennomos.
\end{abstract}

%%%%%%%%%%%%%%%%%%%%%%%%%%%%%%%%%%%%%%%%%%%%%%%%%%%%%%%%%%%%%%%%%%%%%%%%


%%%%%%%%%%%%%%%%%%%%%%%%%%%%%%%%%%%%%%%%%%%%%%%%%%%%%%%%%%%%%%%%%%%%%%%%

\section{Introduction}

Trojans of Jupiter, which reside in the neighbourhood of $L_4$ and $L_5$ Lagrangian points,
serve as an important test of the planetary migration theory
(Morbidelli et al. 2005). Their inclination distribution, namely the large spread of~$I$,
can be explained as a result of chaotic capture during a brief period
when Jupiter and Saturn encountered a 1:2 mean-motion resonance.
Moreover, the Late Heavy Bombardment provides the timing of this resonant
encounter ${\simeq}3.8$\,Gyr ago (Gomes et al. 2005).
It is thus important to understand the population of Trojans accurately.

There are several unresolved problems regarding Trojans, however,
for example the number of families, which is a stringent constraint
for collisional models.
Roig et al. (2008) studied as many as ten suggested families, using relatively
sparse SLOAN data and spectra. They noted most families seem to be heterogeneous
from the spectroscopic point of view, with one exception ---
the C-type Eurybates family. As we argue in this paper, the number
of families (with parent-body size $D \gtrsim 100\,{\rm km}$)
is indeed as low as one.

Another strange fact is the ratio of $L_4$ and $L_5$ Trojans.
Szab\'o et al. (2007) used SLOAN colour data to reach fainter than
orbital catalogues and estimated the ratio to $N(L_4)/N(L_5) = 1.6 \pm 0.1$.
There is no clear explanation for this, since the chaotic capture as
a gravitational interaction should be independent of size or $L_4/L_5$ membership.
Any hypothesis involving collisions would require a relatively recent disruption
of a huge parent body, which is highly unlikely (O'Brien and Morbidelli 2008, D.~O'Brien, personal communication).
This is again related to the actual observed number of Trojan families.

Bro\v z and Vokrouhlick\'y (2008) studied another resonant population,
the so called Hilda group in the 3/2 mean-motion resonance with Jupiter,
and reported only two families: Hilda and Schubart with approximately
200 and 100\,km parent bodies. This number might be in accord
with low collisional probabilities, assuming the Hilda family
is very old and experienced the Late Heavy Bombardment (Bro\v z et al. 2011).

Levison et al. (2009) compared the observed distribution of D-type asteroids
and the model of their delivery from transneptunian region. They found
a good match assuming the D-types (presumably of cometary origin) are
easy-to-disrupt objects (with the strength more than 5 times lower than that of solid ice).
Note that Trojan asteroids are a mixture of C- and D-type objects
and we have to discriminate between them with respect to collisional behaviour.

All of the works mentioned above are a good motivation for us to focus
on asteroid families in the Trojan population. The paper is organised as follows.
First, we describe our data sources and methods in Section~\ref{sec:methods}.
A detailed study of orbital and physical properties of families (and other `false' groupings)
is a matter of Section~\ref{sec:families}. Section~\ref{sec:evolution} is devoted
to the modelling of long-term dynamical evolution.
Finally, there are concluding remarks in Section~\ref{sec:conclusions}.

%%%%%%%%%%%%%%%%%%%%%%%%%%%%%%%%%%%%%%%%%%%%%%%%%%%%%%%%%%%%%%%%%%%%%%
%%%%%%%%%%%%%%%%%%%%%%%%%%%%%%%%%%%%%%%%%%%%%%%%%%%%%%%%%%%%%%%%%%%%%%

\section{Methods}\label{sec:methods}

\subsection{Resonant elements}

We use the symplectic SWIFT integrator (Levison \& Duncan 1994)
for orbital calculations. Our modifications include 
a second order scheme of Laskar \& Robutel (2001)
and on-line digital filters, which enable us to compute
suitable resonant proper elements:
libration amplitude~$d$ of the $a - a'$ oscillations,
where $a$ is the osculating semimajor axis of an asteroid
and $a'$ that of Jupiter,
eccentricity $e$
and inclination $\sin I$.
(In figures, we usually plot a mean value $\bar a$ of semimajor axis
plus the libration amplitude~$d$.)
We employ their definition from Milani (1993).
The source of initial osculating elements is the AstOrb catalogue, version ${\rm JD} = 2455500.5$
(Oct 31st 2010).

There are actually two independent filters running in parallel:
in the first one, we sample osculating elements every 1\,yr,
compute mean elements using the filter sequence B, B with decimation factors 3, 3 (refer to Quinn, Tremaine \& Duncan 1991)
a store this data in a buffer spanning 1\,kyr.
We then estimate the libration frequency~$f$
by a linear fit of $\phi(t) = \lambda - \lambda' - \chi$,
where $\lambda$, $\lambda'$ are the mean longitudes of an asteroid and Jupiter
and $\chi = \pm 60^\circ$ for $L_4$ or $L_5$ respectively.
The revolution of angle $\phi(t)$ must not be confined to the interval $[0, 360^\circ)$, of course.
The amplitude of~$d$ is computed for the already known~$f$ by a~discrete Fourier transform.
Finally, an off-line running-average filter with a window 1\,Myr is used to smooth the data.%
\footnote{Equivalently, we may compute the amplitude~$D$ of mean longitudes $\lambda-\lambda'$.
Anyway, there is a linear relation between $d$ and $D$.}

In the second filter, we compute proper eccentricity~$e$ and proper inclination~$\sin I$ by
sampling osculating elements (1\,yr step),
computing mean elements using a filter sequence A, A, B and decimation factors 10, 10, 3,
and then we apply a frequency modified Fourier transform (Nesvorn\'y \& \v Sidlichovsk\'y 1997),
which gives us the relevant proper amplitudes.

\begin{figure}
\centering
\includegraphics[width=6.0cm]{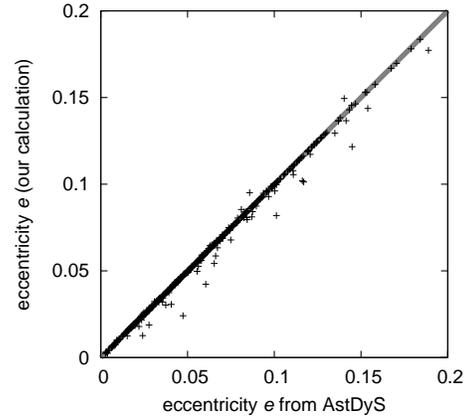}
\caption{Comparison of the resonant eccentricity calculated by our code
to that of Kne\v zevi\'c \& Milani (AstDyS catalogue).
There is a line $x=y$ to aid a comparison.}
\label{trojans-hcm_20100915_follow.proper.out_dfD_0Myr_astdys4_e}
\end{figure}

The values of the resonant elements agree very well with those listed
in the AstDyS catalogue by Kne\v zevi\'c \& Milani
(see Figure~\ref{trojans-hcm_20100915_follow.proper.out_dfD_0Myr_astdys4_e}).
There are only few outliers, probably due to a different time span of integration.
We computed proper elements for 2647 $L_4$ and 1496 $L_5$ Trojan asteroids.%
\footnote{The data are available in an electronic form on our web site
{\tt http://sirrah.troja.mff.cuni.cz/\~{}mira/mp/}.
We use also one-apparition orbits for the purposes of physical studies.
Of course, orbital studies require more precise multi-apparition data.}
This sample is roughly twice larger than previously analysed.
The ratio of populations valid for $H \lesssim 15\,{\rm mag}$
asteroids is thus $N(L_4)/N(L_5) \simeq 1.8$.

\begin{figure*}
\centering
\begin{tabular}{cc}
$L_4$ Trojans & $L_5$ Trojans \\
\includegraphics[width=7cm]{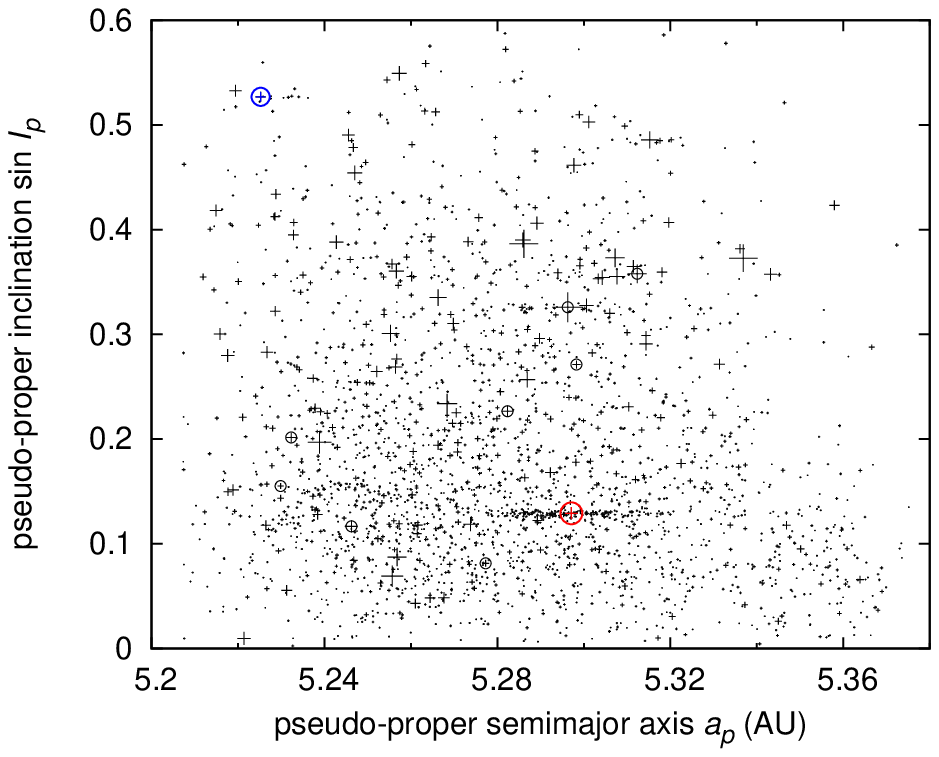} & 
\includegraphics[width=7cm]{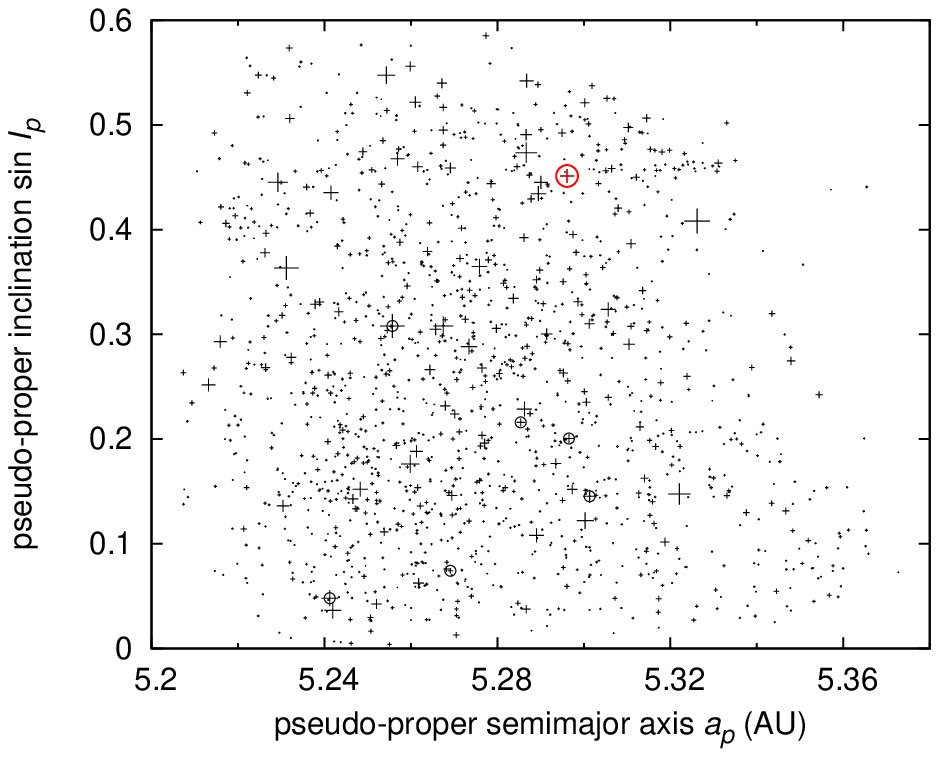} \\
\includegraphics[width=7cm]{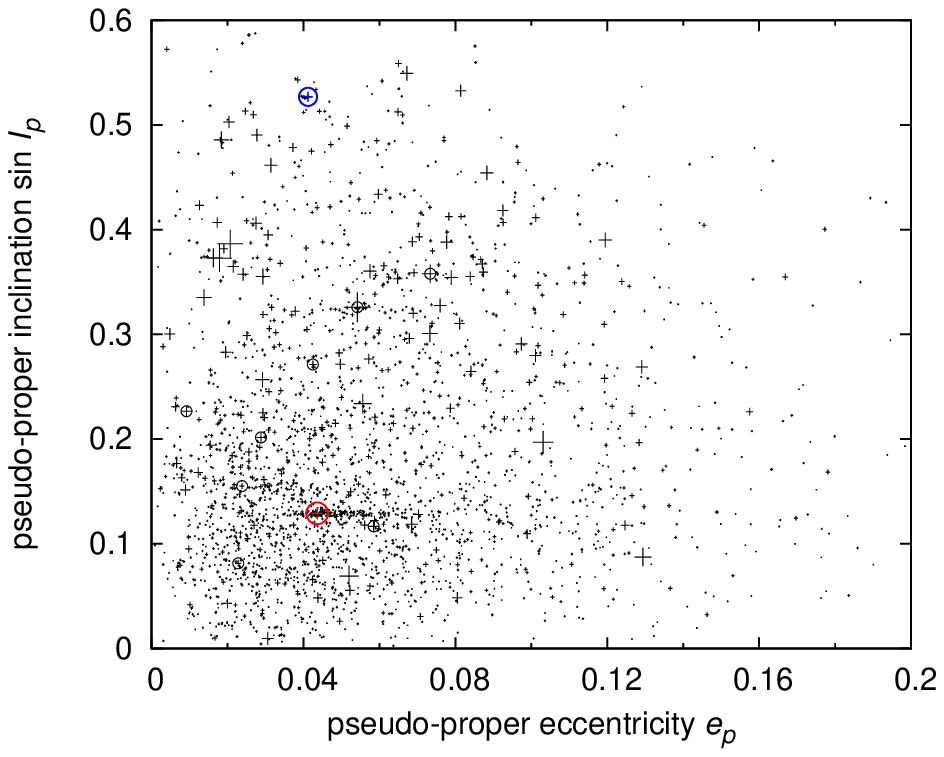} & 
\includegraphics[width=7cm]{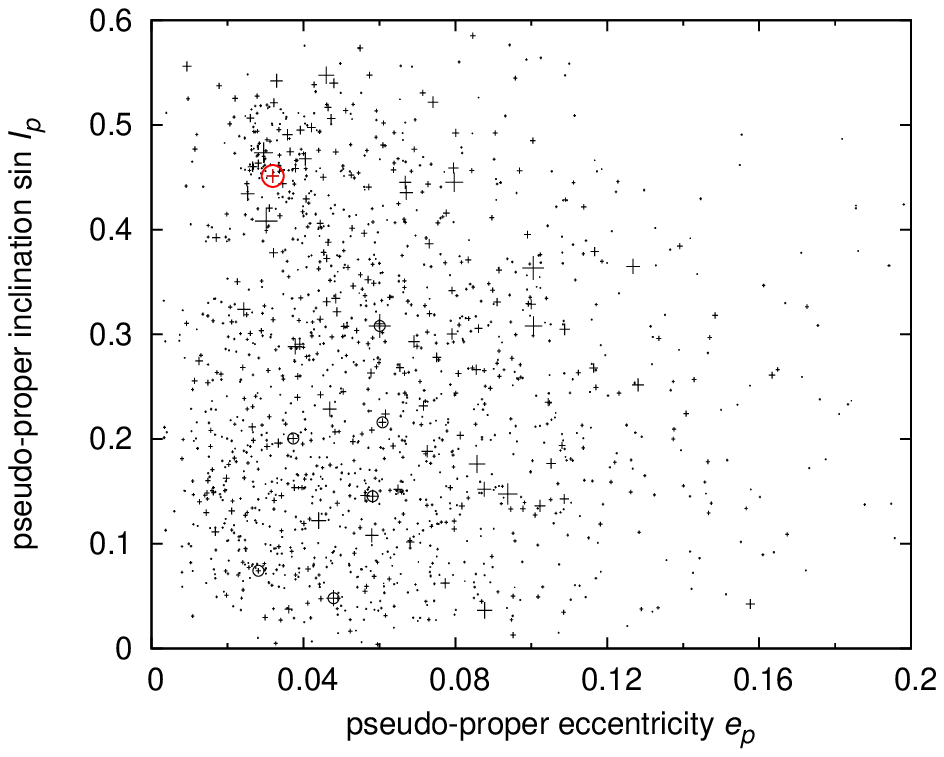} \\
\end{tabular}
\caption{The resonant elements $(a \equiv \bar a + d, \sin I)$ and $(e, \sin I)$ for $L_4$ and $L_5$ Trojans.
The crosses indicate relative sizes of bodies, taken either from
the AstOrb catalogue or computed from absolute magnitude~$H$ and geometric albedo~$p_V$.
In this plot, we assumed $p_V = 0.058$ for $L_4$ Trojans and $0.045$ for those in $L_5$
(it corresponds to medians of known $p_V$'s).
The asteroids (3548)~Eurybates in $L_4$ and (4709)~Ennomos in $L_5$,
around which significant clusters are visible, are shown in red.
Moreover, the asteroid (9799)~1996~RJ in $L_4$, which is surrounded
by a small cluster, is denoted by a blue circle. (This cluster is so tight,
that its members are located inside the circle on the $(e, \sin I)$ plot.)}
\label{trojans_L4_arer_sizes}
\end{figure*}

The overall distribution of Trojans in the $(d, e, \sin I)$ space
is shown in Figure~\ref{trojans_L4_arer_sizes}. Note there is only one
cluster visible immediately in the bottom-left panel --- around (3548)~Eurybates.
The reason is its tight confinement in inclinations ($\sin I = 0.125\hbox{ to }0.135$).

%%%%%%%%%%%%%%%%%%%%%%%%%%%%%%%%%%%%%%%%%%%%%%%%%%%%%%%%%%%%%%%%%%%%%%

\subsection{Hierarchical clustering}

\begin{figure*}
\centering
\includegraphics[width=5.2cm]{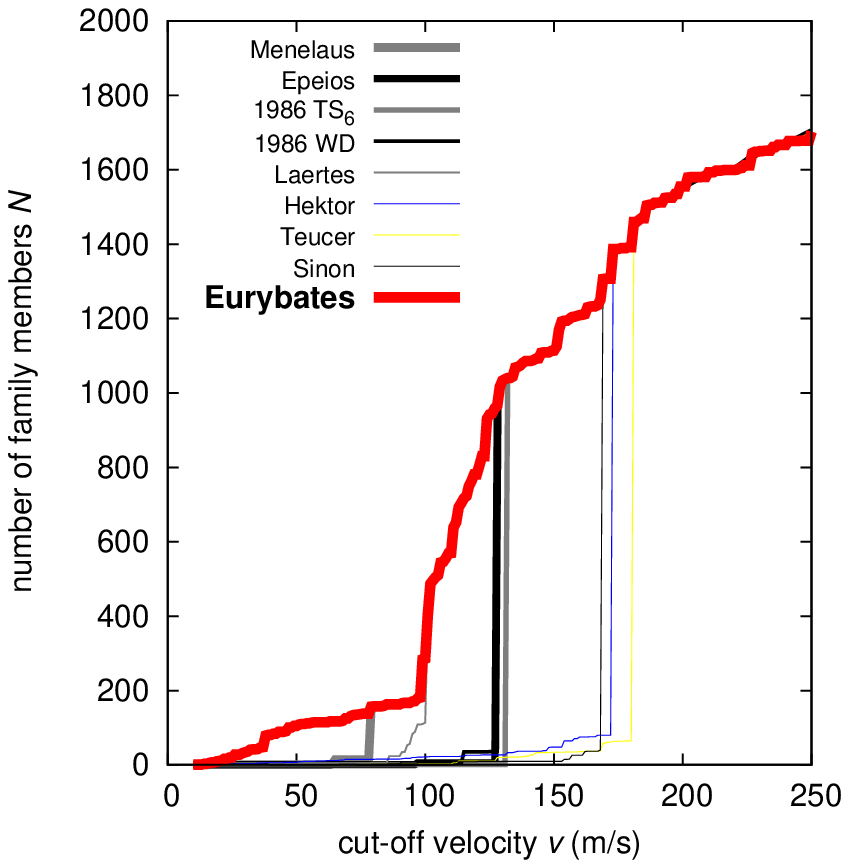}
\includegraphics[width=5.2cm]{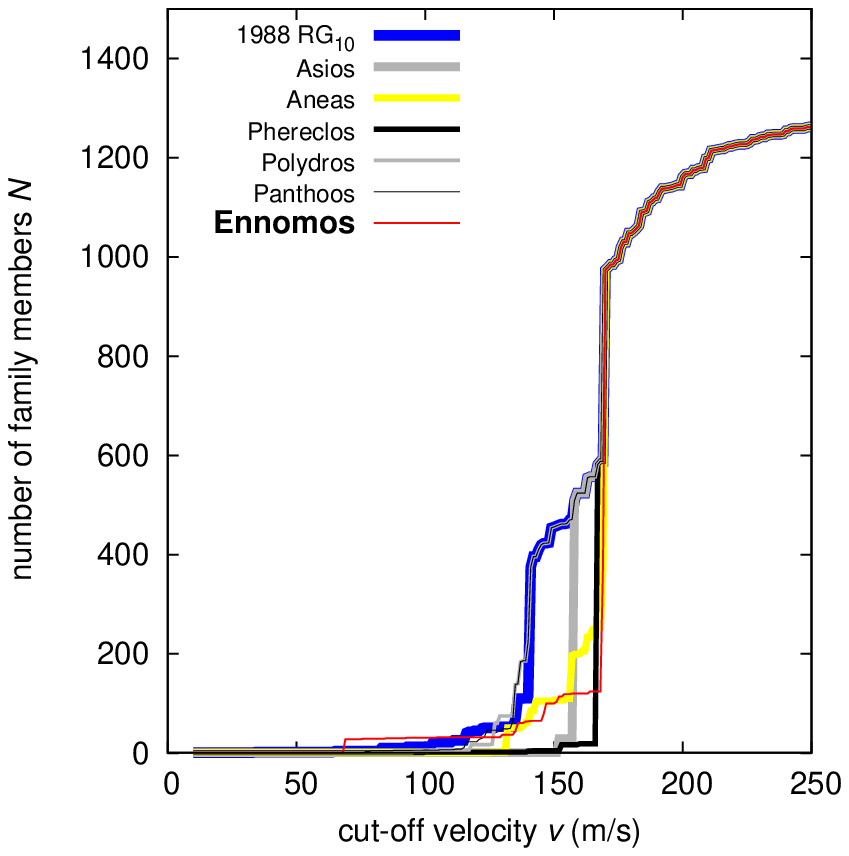}
\includegraphics[width=5.2cm]{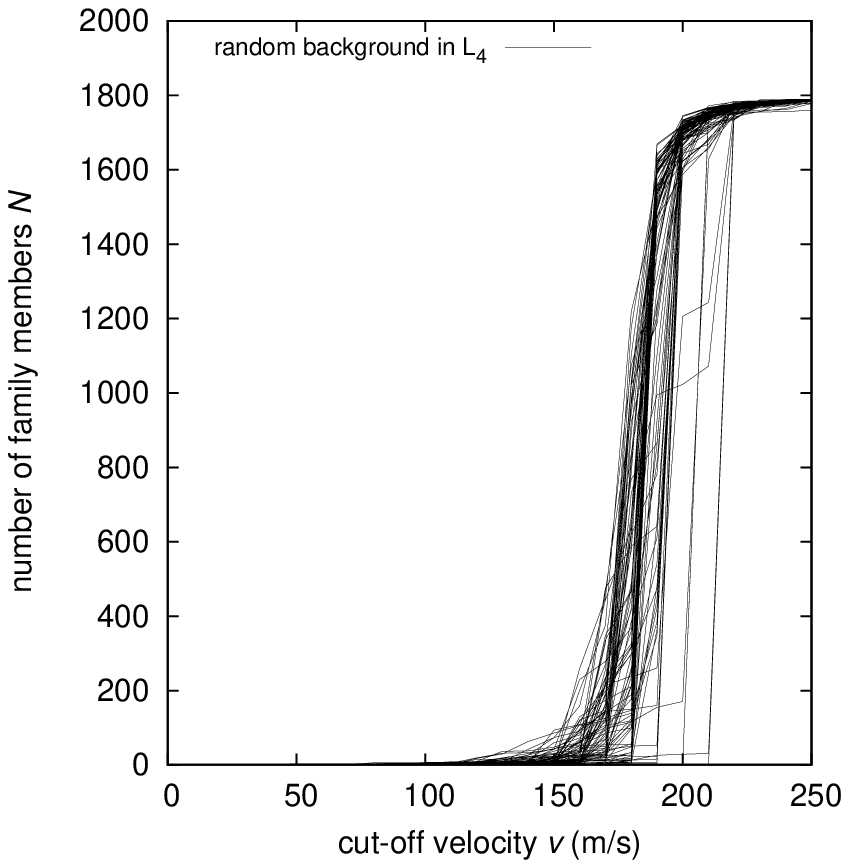}
\caption{Left panel: The dependence of the number of family members $N$ on the cut--off velocity $v_{\rm cutoff}$
computed by the hierarchical clustering method. Only clusters among $L_4$ Trojans are included in this plot.
Middle panel: The same $N(v_{\rm cutoff})$ dependence for $L_5$ Trojans.
Right panel: Artificial clusters selected from {\em random\/} distribution of asteroids
generated in the same volume of the $(d, e, \sin I)$ space.}
\label{L4_Nv_random}
\end{figure*}

In order to detect clusters in the resonant element space
we use a hierarchical clustering method (Zappal\'a et al. 1994)
with a standard metric $d_1$, with $\delta a$ substituted by~$d$.
We run the HCM code many times with various starting bodies
and different cut--off velocities $v_{\rm cutoff}$
and determine the number of bodies~$N$ in the given cluster.
We find the $N(v_{\rm cutoff})$ dependence a very useful diagnostic tool.
We can see these dependences for $L_4$ and $L_5$ Trojans in Figure~\ref{L4_Nv_random}.

It is easy to recognise, if a cluster has a concentration towards the centre
--- even at low $v_{\rm cutoff}$ it must have more than one member ($N\gg1$).
It is also instructive to compare clusters with a random background (thin lines),
which we generated artificially by a random-number generator in the same volume
of the $(d, e, I)$ space. Insignificant (random) clusters usually exhibit an abrupt
increase of $N$ at a high cut--off velocity.

As starting bodies we selected those listed in Roig et al. (2008).
Only three clusters, namely the Eurybates, Aneas, 1988~RG$_{10}$
seem to be somewhat concentrated, i.e., denser than the background.
The Hektor cluster is also concentrated but it contains only
a relatively low number of members (20 to 70) before it merges with the background.
In other words, smaller asteroids do not seem concentrated around (624)~Hektor.
Remaining clusters are more or less comparable to the background.

Nevertheless, we report a detection of a previously unknown cluster
around (4709)~Ennomos in $L_5$. It is relatively compact, since the minimum
cut-off velocity is 70\,m/s only. The cluster contains mostly
small bodies which were discovered only recently.

Finally, let us point out a very tight cluster around (9799) 1996~RJ,
associated already at $v_{\rm cutoff} = 20\,{\rm m}/{\rm s}$.
It is located at high inclinations and contains 9~bodies, three of them having short arcs.
The cluster seems peculiar in the osculating element space too
since it exhibits a non-random distribution of nodes and perihelia (see Table~\ref{9799_tab}).
This is similar to very young families like the Datura (Nesvorn\'y et al. 2006)
and it makes the 1996~RJ cluster a particularly interesting case with respect
to collisional dynamics. Because one has to use slightly different methods for studies
of such young families we postpone a detailed analysis to a next paper.

\begin{table*}
\caption{List of nine members of the (9799) 1996~RJ cluster and their proper
$(a, e, \sin I)$ and osculating $(\Omega_{\rm osc}, \varpi_{\rm osc})$ elements
and absolute magnitude~$H$.
Note the distribution of nodes and perihelia is not entirely uniform.
Asteroids with short-arc orbits (${<}60$~days) are denoted by a * symbol.}
\begin{tabular}{rlcrrrrrr}
\hline
number & designation & & \multicolumn{1}{c}{$a$} & \multicolumn{1}{c}{$e$} & \multicolumn{1}{c}{$\sin I$} & $\Omega_{\rm osc}$ & $\varpi_{\rm osc}$ & $H / {\rm mag}$ \\
\hline
  9799 & 1996 RJ         &   & 5.2252 & 0.0412 & 0.5269 & 115.4 & 259.6 &  9.9 \\
 89938 & 2002 FR$_{4}$   &   & 5.2324 & 0.0394 & 0.5274 &  70.0 &  23.1 & 12.5 \\
226027 & 2002 EK$_{127}$ &   & 5.2316 & 0.0399 & 0.5263 &  62.8 & 352.9 & 12.6 \\
243316 & 2008 RL$_{32}$  &   & 5.2340 & 0.0398 & 0.5268 &  27.3 & 358.2 & 12.8 \\
       & 2005 MG$_{24}$  &   & 5.2275 & 0.0404 & 0.5252 & 172.3 & 236.5 & 13.1 \\
       & 2008 OW$_{22}$  & * & 5.2276 & 0.0401 & 0.5274 &  53.7 & 340.9 & 13.9 \\
       & 2009 RA$_{17}$  & * & 5.2258 & 0.0409 & 0.5272 & 257.7 & 194.5 & 13.7 \\
       & 2009 RK$_{63}$  & * & 5.2305 & 0.0407 & 0.5260 &  56.4 &   5.6 & 12.8 \\
       & 2009 SR$_{30}$  &   & 5.2362 & 0.0409 & 0.5258 & 103.6 &  22.0 & 13.3 \\
\hline
\end{tabular}
\label{9799_tab}
\end{table*}

\begin{figure}
\centering
\includegraphics[width=8cm]{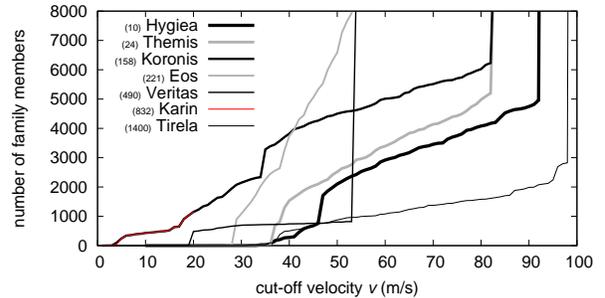}
\caption{The $N(v_{\rm cutoff})$ dependence for seven outer main-belt families.
If we would consider only a subset of asteroids brighter than $H = 15\,{\rm mag}$, which is an approximate
observational limit for Trojans, the $N(v_{\rm cutoff})$ dependencies would be qualitatively the same,
only slightly shifted to larger cut--off velocities.}
\label{main_belt_families_Nv_ALL}
\end{figure}

Let us compare Trojan clusters to the well known asteroid families
in the outer Main Belt (Figure~\ref{main_belt_families_Nv_ALL}).
Most families (e.g., Themis, Koronis, Eos) exhibit a steady
increase of~$N$ until they merge with another families or the entire outer Main Belt.
Eurybates, Aneas and 1988~RG$_{10}$ are the only Trojan clusters which behave in a similar fashion.
The Veritas family (dynamically young, Nesvorn\'y et al. 2003) exhibits a different behaviour
--- for a large interval of $v_{\rm cutoff}$ the number of members~$N$ remains
almost the same, which indicates a clear separation from the background population.
With respect to the $N(v_{\rm cutoff})$ dependence, the Ennomos cluster is similar
to Veritas.

%%%%%%%%%%%%%%%%%%%%%%%%%%%%%%%%%%%%%%%%%%%%%%%%%%%%%%%%%%%%%%%%%%%%%%

\subsection{Size-frequency distribution}

\begin{figure*}
\centering
\includegraphics[width=7cm]{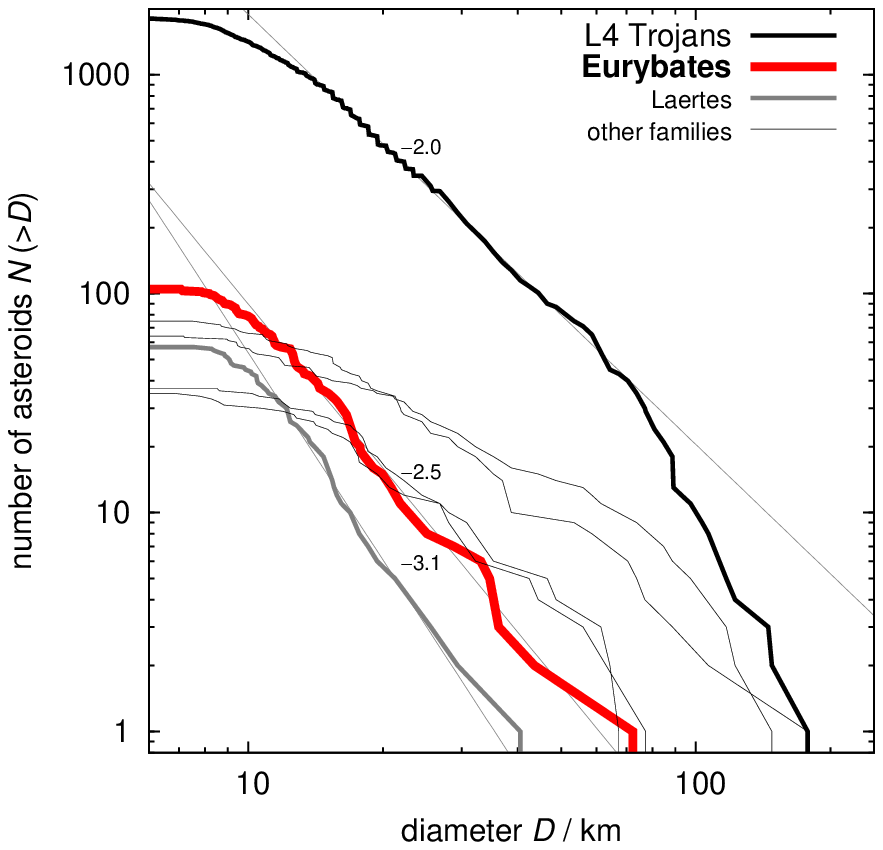}
\includegraphics[width=7cm]{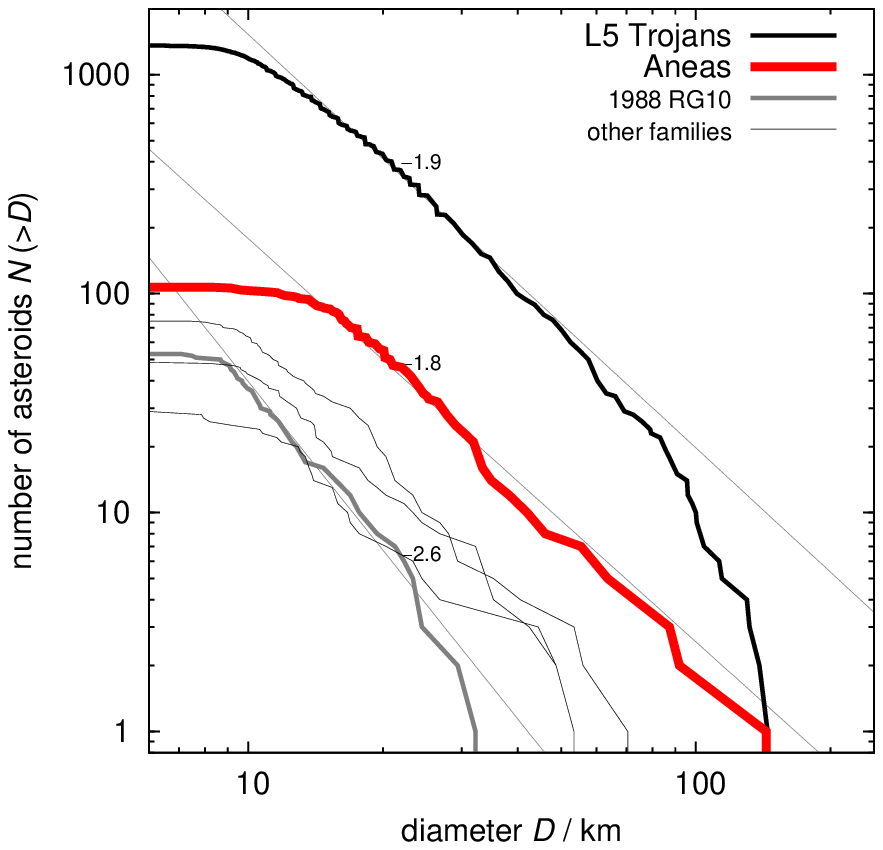}
\caption{Left panel: size distributions of $L_4$ Trojans
and the following clusters (there is a selected cut--off velocity in the parenthesis):
Eurybates ($v_{\rm cutoff} = 50\,{\rm m}/{\rm s}$),
Laertes (94),
Hektor (160),
Teucer (175),
Sinon (163),
1986~WD (120).
Right panel: SFD's of $L_5$ Trojans and the following clusters:
1988~RG$_{10}$ (at $v_{\rm cutoff} = 130\,{\rm m}/{\rm s}$),
Aneas (150),
Asios (155),
Panthoos (130),
Polydoros (130).}
\label{L4_size_distribution}
\end{figure*}

At first, let us assume a single value of albedo for all family members.
It is a reasonable assumption if the family is of collisional origin.
We can then calculate sizes from absolute magnitudes and construct
size-frequency distributions. Figure~\ref{L4_size_distribution}
shows a comparison of SFD's for the clusters detected by the HCM%
\footnote{Of course, we have to select a `suitable' value of the cut--off velocity
for all clusters. Usually, we select that value for which $N(v_{\rm cutoff})$ is flat.
Size-frequency distribution is not very sensitive to this selection anyway.}
and for the whole population of $L_4$ and $L_5$ Trojans.

A slope~$\gamma$ of the cumulative distribution $N({>}D) \propto D^{\gamma}$ is an indicative parameter.
For $L_4$ and $L_5$ Trojans it equals to $-2.0\pm 0.1$ and $-1.9\pm0.1$ in the intermediate
size range 15 to 60\,km. (These numbers are compatible with the study of Yoshida \& Nakamura 2008.)
The slope is steeper at large sizes.
The uncertainties are mainly due to a freedom in selection of the size range
and the difference between $L_4$ and $L_5$ SFD's does not seem significant.
The clusters have typically similar slope as background (within 0.1 uncertainty),
thought sometimes the results are inconclusive due to small
number of members. The Eurybates family with $-2.5\pm0.1$ slope
is on the other hand significantly steeper than the mean slope
of the whole Trojan population.%
\footnote{Thought the number of the Eurybates members (105)
is so small that it almost does not affect the mean slope
of the whole $L_4$ population.}
There are two more groups which exhibit a relatively steep slope,
namely Laertes in $L_4$ ($\gamma = -3.1$) and 1988~RG$_{10}$ in $L_5$ ($\gamma = -2.6$).

\begin{figure}
\centering
\includegraphics[width=8cm]{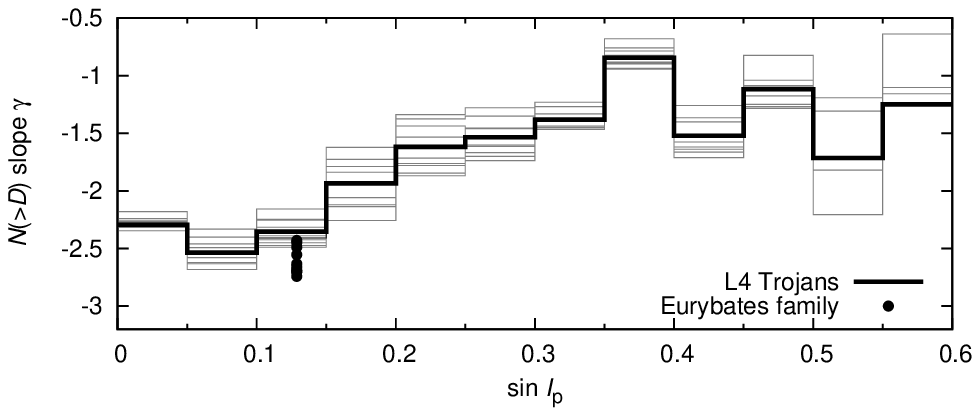}
\includegraphics[width=8cm]{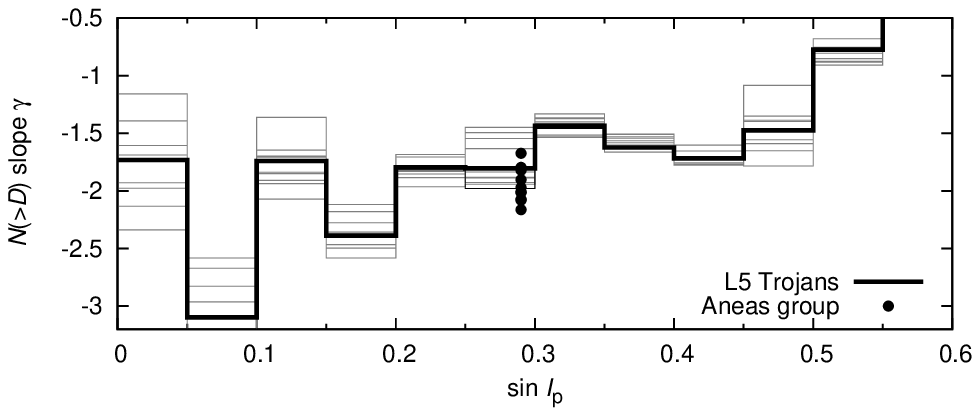}
\caption{Slopes~$\gamma$ of the size-frequency distributions $N({>}D)$ for $L_4$ and $L_5$ Trojans
and their dependence on the inclination~$\sin I$.
The range of diameters for which the SFD's were fitted is $D_{\rm min} = 12\,{\rm km}$, $D_{\rm max} = 30\,{\rm km}$.
Thin lines were calculated for different ranges, which were varied
as $D_{\rm min} \in (10, 15)\,{\rm km}$, $D_{\rm max} \in (20, 40)\,{\rm km}$.
Their spread indicates the uncertainty of~$\gamma$ in a given interval of~$\sin I$.
The populations are observationally complete down to $D \simeq 10\,{\rm km}$,
because the characteristic change of slope due to incompleteness
occurs at smaller sizes (see also Yoshida and Nakamura 2008).}
\label{trojans-hcm_20090623_trojans_SINI.lsm_ALL}
\end{figure}

We should be aware, however, that even the background exhibits
a trend with respect to inclinations (see Figure~\ref{trojans-hcm_20090623_trojans_SINI.lsm_ALL}).
Slope~$\gamma$ typically decreases with inclination~$\sin I$,
which is especially prominent in case of the $L_4$~cloud.
We have to admit if we compare the Eurybates family to its surroundings only
($\sin I = 0.1\hbox{ to }0.15$), the difference in slopes is not so prominent.
An interesting feature of the $L_5$~cloud is a dip in the interval
$\sin I = 0.05\hbox{ to }0.1$. This corresponds to the
approximate location of the 1988~RG$_{10}$ group.

The $\gamma(\sin I)$ dependence among Trojans is not unique.
E.g. low-inclination bodies in the J3/2 resonance also have the SFD
steeper than background ($\gamma = -2.5\pm0.1$ versus $-1.7\pm0.1$),
without any clear family and a few big interlopers.
May be, this feature reflects different {\em source reservoirs\/}
of low- and high-inclination bodies among Trojans and J3/2?%
\footnote{Both Trojan and J3/2 regions are dynamically unstable
during Jupiter--Saturn 1:2 mean motion resonance,
so we expect the same bodies entering Trojan region may enter J3/2.}
It may be also in concert with a colour--inclination dependence
reported by Szab\'o et al. (2007).

We also test albedo distributions dependent on size,
since the measurements by Fern\'andez et al. (2009) suggested
small Trojans are significantly brighter and thus smaller.
Large asteroids have $p_V = 0.044\pm0.008$ while small $p_V = 0.12\pm0.06$.
This is a significant change of the SFD, which occurs around the size $D \simeq 30\,{\rm km}$.
The SFD thus becomes shallower below this size,
e.g. for Eurybates we would have $\gamma = -1.6$ and for L4 Trojans $\gamma = -1.5$,
so the SFD's become comparable with respect to the slope.
Thought, as we stated above, for a real collisional family
we expect the albedo distribution to be rather homogeneous and independent of size.

%%%%%%%%%%%%%%%%%%%%%%%%%%%%%%%%%%%%%%%%%%%%%%%%%%%%%%%%%%%%%%%%%%%%%%

\subsection{Colour and spectral data}

We used the Sloan Digital Sky Survey Moving Object catalogue version 4 (SDSS-MOC4) to check the
families are spectrally homogeneous, as we expect. Due to a larger uncertainty in the $u$ colour
in SDSS-MOC4, we used the color indices $a^*$ and $i-z$, where $a^* = 0.89(g-r) + 0.45(r-i) - 0.57$
(defined by Parker et al. 2008).

\begin{figure}
\centering
\includegraphics[width=4cm]{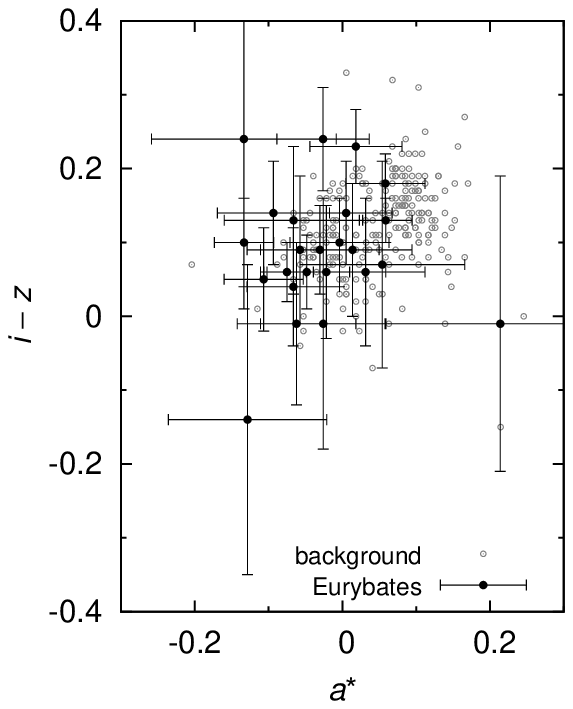}
\includegraphics[width=4cm]{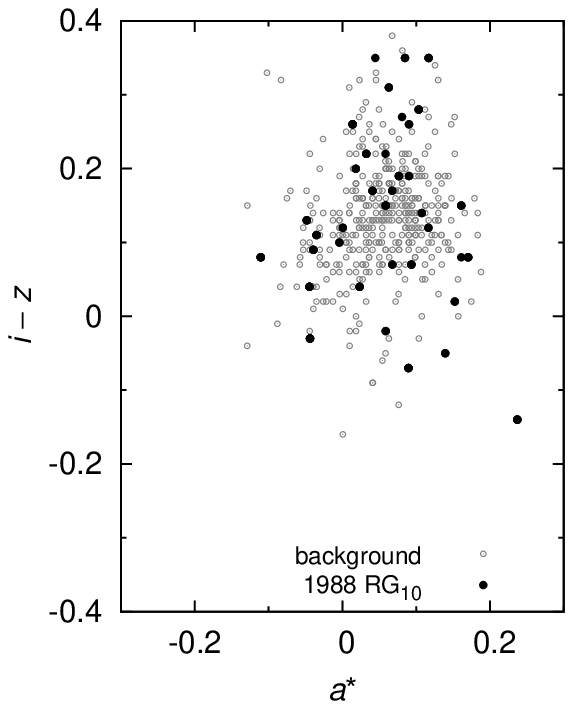}
\caption{Left panel: The $(a^*, i-z)$ colours for the L$_4$ Trojans (gray dots) and the Eurybates family (black dots with error bars).
The distributions differ significantly in this case.
Right panel: A similar comparison for the L$_5$ Trojans and the 1988~RG$_{10}$ group,
which seem to be indistinguishable.}
\label{pc1_vs_pc2_Err_Eurybates}
\end{figure}

The result is shown in Figure~\ref{pc1_vs_pc2_Err_Eurybates}. It is clearly visible that the distribution
of the Eurybates family in the space of $(a^*, i-z)$ colours is different from the Trojan background.
On contrary, the 1988 RG$_{10}$ group covers essentially the same area as the background.
The Aneas is only slightly shifted towards larger $a^*$ and $i-z$ with respect to the background.
There is a lack of data for the Ennomos group --- three bodies are not sufficient to compare
the colour distributions.

Alternatively, we may use principal component analysis of the SDSS colour indices.
We use only data with uncertainties smaller than 0.2\,mag,
which resulted in 70\,887 records.
We calculated eigenvalues 
($\lambda_{1,2,3,4} = 0.173$,
0.0532,
0.0249,
0.0095),
corresponding eigenvectors and constructed the following three principal components (Trojanov\'a 2010):
\begin{eqnarray}
{\rm PC}_1 &=&     0.235\, (u-g)    +0.416\, (g-r)    +0.598\, (g-i) \nonumber \\
           & & +\, 0.643\, (g-z)                                     \,, \\
{\rm PC}_2 &=&     0.968\, (u-g)    -0.173\, (g-r)    -0.147\, (g-i) \nonumber \\
           & & -\, 0.106\, (g-z)                                     \,, \\
{\rm PC}_3 &=&     0.078\, (u-g)    +0.601\, (g-r)    +0.330\, (g-i) \nonumber \\
           & & -\, 0.724\, (g-z)                                     \,,
\end{eqnarray}
which have a clear physical interpretation:
PC$_1$ corresponds to an overall slope,
PC$_2$ is a variability in the $u$~band, and
PC$_3$ a depth of the $1\,\mu{\rm m}$ absorption band.
The Eurybates family is different from Trojans in all three principal components
(mean PC$_1$ of the Eurybates members is smaller, PC$_2$ and PC$_3$ larger).
The Aneas group has the same distribution of PC$_2$ and PC$_3$ as Trojans
and the 1988~RG$_{10}$ group is similar to Trojans even in all three components.

Hence, we confirm the Eurybates family seems distinct in color even in the fourth version
of the SDSS-MOC. This fact is consistent with previous work of Roig et al. (2008),
who used third version of the same catalogue and classified Eurybates family members
as C-type asteroids.

Finally, note that De Luise et al. (2010) pointed out an absence of ice spectral
features at 1.5 and $2.0\,\mu{\rm m}$ on several Eurybates members
and Yang and Jewitt (2007) concluded the same for (4709) Ennomos.
This puzzling fact may indicate that pure ice covers at most
10\,\% of the Ennomos surface.

%We subtracted solar analogue:
%$(u-g)_\odot = 1.32$,
%$(g-r)_\odot = 0.45$,
%$(r-i)_\odot = 0.10$,
%$(i-z)_\odot = 0.04$.

% # vlastni cisla matice a():
%  0.17270036
%  5.31925820E-02
%  2.49198601E-02
%  9.46618244E-03

% # vlastni vektory matice a() v radcich:
%      0.23546046      0.41565195      0.59843224      0.64317226
%      0.96814215     -0.17290485     -0.14676926     -0.10612940
%      0.07798612      0.60094780      0.32988197     -0.72384918
%     -0.03431175     -0.66045374      0.71520305     -0.22607116

%%%%%%%%%%%%%%%%%%%%%%%%%%%%%%%%%%%%%%%%%%%%%%%%%%%%%%%%%%%%%%%%%%%%%%

\subsection{Impact disruption model}\label{sec:disruption}

We use a simple model of an isotropic disruption from
Farinella et al. (1994). The distribution of velocities
"at infinity" follows the function
\begin{equation}
{\rm d}N(v) = C v (v^2 + v_{\rm esc}^2)^{-(\alpha+1)/2}\,,\label{dN_v}
\end{equation}
with the exponent $\alpha$ being a~free parameter,
$C$ a normalisation constant
and $v_{\rm esc}$ the escape velocity from the parent body,
which is determined by its size~$R_{\rm PB}$ and mean density~$\rho_{\rm PB}$.
The distribution is cut at a selected maximum allowed 
velocity $v_{\rm max}$ to prevent outliers. We typically use $v_{\rm max} = 300\,{\rm m}/{\rm s}$.
The orientations of velocity vectors in space are assigned randomly.
We assume the velocity of fragments is independent on their size.%
\footnote{If we use a size-dependent relation for velocities similar
to Vokrouhlick\'y et al. (2006), our results do not change much, because
the overall shape of the velocity distribution is quite similar
to the size-independent case.}

There are several more free parameters, which determine the initial
shape of the family in the space of proper elements:
initial osculating eccentricity~$e_{\rm i}$ of the parent body,
initial inclination~$i_{\rm i}$,
as well as true anomaly~$f_{\rm imp}$
and argument of perihelion~$\omega_{\rm imp}$
at the time of impact disruption.

\begin{figure*}
\centering
\includegraphics[width=16cm]{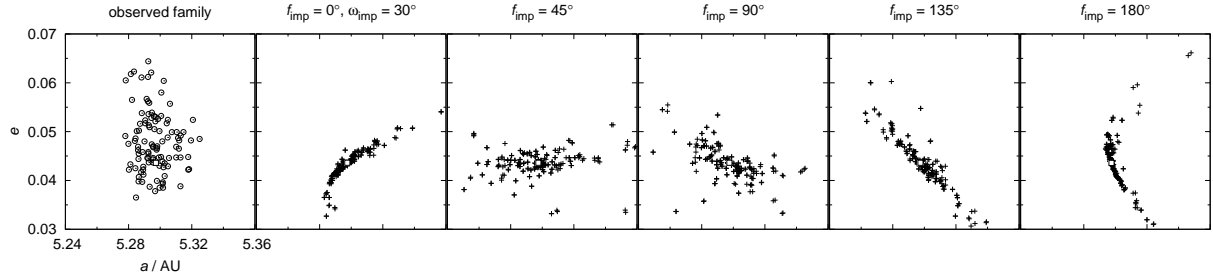}%
\caption{A comparison between the observed Eurybates family (open circles) and synthetic families (crosses)
just after the impact disruption computed for
several values of $f_{\rm imp} = 0^\circ, 45^\circ, 90^\circ, 135^\circ, 180^\circ$
and
$\omega_{\rm imp} = 30^\circ$,
$R_{\rm PB} = 47\,{\rm km}$,
$\rho_{\rm PB} = 1300\,{\rm kg}/{\rm m^3}$.
Different geometry in $f$, $\omega$ produces a slightly different
cluster, nevertheless, it is always tighter than the observed family.
The position of the asteroid (3548)~Eurybates is denoted by a square.}
\label{eurybates-impact4_f0_ae}
\end{figure*}

An example of a synthetic family just after disruption and its comparison
to the observed Eurybates family is shown in Figure~\ref{eurybates-impact4_f0_ae}.
Usually, there is a significant disagreement between this
simple model of impact disruption and the observations.
Synthetic families usually look like thin `filaments' in the $(d, e, \sin I)$ space,
which are curved due to the mapping from osculating elements to resonant ones.
On the other hand, observed groups among Trojans are much more spread.
However, this only indicates an importance of further long-term evolution
by chaotic diffusion and possibly by planetary migration.%
\footnote{Only very young clusters like the Karin family (Nesvorn\'y et al. 2002)
exhibit this kind of a `filament' shape.}

In case of the Ennomos group members, they are distributed mostly
on larger semimajor axes than (4709) Ennomos, thought isotropic
impact disruptions produce fragments distributed evenly on larger
and smaller~$a$. May be, it is an indication of an anisotropic
velocity field? Or a different parent body of this cluster?

%%%%%%%%%%%%%%%%%%%%%%%%%%%%%%%%%%%%%%%%%%%%%%%%%%%%%%%%%%%%%%%%%%%%%%

\subsection{Planetary migration}

If asteroid families are very old, planetary migration might influence
their current shape. In order to study of late stages of planetary migration,
which is caused by interactions with a planetesimal disk, we construct the following model.
We treat the migration analytically within a modified version of the numerical
symplectic SWIFT-RMVS3 integrator (Levison \& Duncan 1994), which accounts
for gravitational perturbations of the Sun and four giant planets
and includes also an energy-dissipation term, as described in Bro\v z et al. (2011).
The speed of migration is characterised by the exponential time scale $\tau_{\rm mig}$
and the required total change of semimajor axis $a_{\rm i} - a_{\rm f}$.
We use an eccentricity damping formula too, which simulates
the effects of dynamical friction and prevent an unrealistic increase of eccentricities (Morbidelli et al. 2010).
The amount of damping is determined by the parameter $e_{\rm damp}$.

We try to adjust initial orbital parameters of planets and the parameters
of migration in such a way to end up at currently observed orbits.
The integration time step is $\Delta t = 36.525$~days
and the time span is usually equal to $3 \tau_{\rm mig}$,
when planetary orbits almost stop to migrate.

%%%%%%%%%%%%%%%%%%%%%%%%%%%%%%%%%%%%%%%%%%%%%%%%%%%%%%%%%%%%%%%%%%%%%%

\subsection{Inefficient Yarkovsky/YORP effect}

On long time scales, the Yarkovsky thermal force might cause significant
perturbations of orbits. We use an implementation of the Yarkovsky thermal effect
in the SWIFT N-body integrator (Bro\v z 2006). It includes both the diurnal
and seasonal variants.

The YORP effect (thermal torques affecting spin states; Vokrouhlick\'y et al. 2006)
was not taken into account in our simulations. The reason is that the respective
time scale~$\tau_{\rm YORP}$ is of the order 100\,My to 1\,Gyr
So as a `zero' approximation, we neglect the YORP effect on these "short"
time scales and keep the orientations of the spin axes fixed.

For Trojan asteroids captured in a zero-order mean motion resonance, the Yarkovsky
perturbation only affects the position of libration centre (Moldovan et al. 2010).
Note that the perturbation acts `instantly' --- there is no systematic secular drift
in eccentricity nor in other proper elements which is an important difference
from first-order resonances, where a $e$-drift is expected (Bro\v z \& Vokrouhlick\'y 2008, Appendix~A).
This is another reason we do not need a detailed YORP model here.

The thermal parameter we use are reasonable estimates for C/X-type bodies:
$\rho_{\rm surf} = \rho_{\rm bulk} = 1300\,{\rm kg}/{\rm m}^3$ for the surface and bulk densities,
$K = 0.01\,{\rm W}/{\rm m}/{\rm K}$ for the surface thermal conductivity,
$C = 680\,{\rm J}/{\rm kg}$ for the heat capacity,
$A = 0.02$ for the Bond albedo and
$\epsilon_{\rm IR} = 0.95$ for the thermal emissivity.

%%%%%%%%%%%%%%%%%%%%%%%%%%%%%%%%%%%%%%%%%%%%%%%%%%%%%%%%%%%%%%%%%%%%%%
%%%%%%%%%%%%%%%%%%%%%%%%%%%%%%%%%%%%%%%%%%%%%%%%%%%%%%%%%%%%%%%%%%%%%%

\section{Asteroid families and insignificant groups}\label{sec:families}

In this section, we briefly discuss properties
of selected clusters: Eurybates, Ennomos, Aneas and 1988~RG$_{10}$.
We focus on these four clusters, since they seem
most prominent according to our previous analysis.

\subsection{Eurybates family}

The Eurybates family can be detected by the hierarchical clustering method for cut--off velocities
$v_{\rm cutoff} = 38\hbox{ to }78\,{\rm m}/{\rm s}$,
when it merges with Menelaus (see Figure~\ref{L4_Nv_random}).
Yet, we do not rely just on the HCM! Another selection criterion we use
is a `meaningful' shape of the family and its changes with respect to $v_{\rm cutoff}$.
A very important characteristic of the Eurybates family at low values of $v_{\rm cutoff}$
is a tight confinement of inclinations ($\sin I$ within 0.01).
It breaks down at $v_{\rm cutoff} \simeq 68\,{\rm m}/{\rm s}$,
so we consider this value as an upper limit.
The Eurybates family is also confined in semimajor axis,
being approximately twice smaller than other groups.

The diameter of the parent body is $D_{\rm PB} \doteq 97\,{\rm km}$ for albedo $p_V = 0.055$
if we sum the volumes of the known bodies. Of course, in reality it is slightly larger
due to observational incompleteness. If we prolong the slope of the SFD $\gamma = -2.5$
down to zero we obtain $D_{\rm PB} \doteq 110\,{\rm km}$.
The geometric method of Tanga et al. (1999) gives an upper limit
$D_{\rm PB} \simeq 130\,{\rm km}$.

Spectral slopes of family members are rather homogeneous and correspond
to C/P-types (Roig et al. 2008).

%%%%%%%%%%%%%%%%%%%%%%%%%%%%%%%%%%%%%%%%%%%%%%%%%%%%%%%%%%%%%%%%%%%%%%

\subsection{Ennomos group}

\begin{figure*}
\centering
\includegraphics[width=6cm]{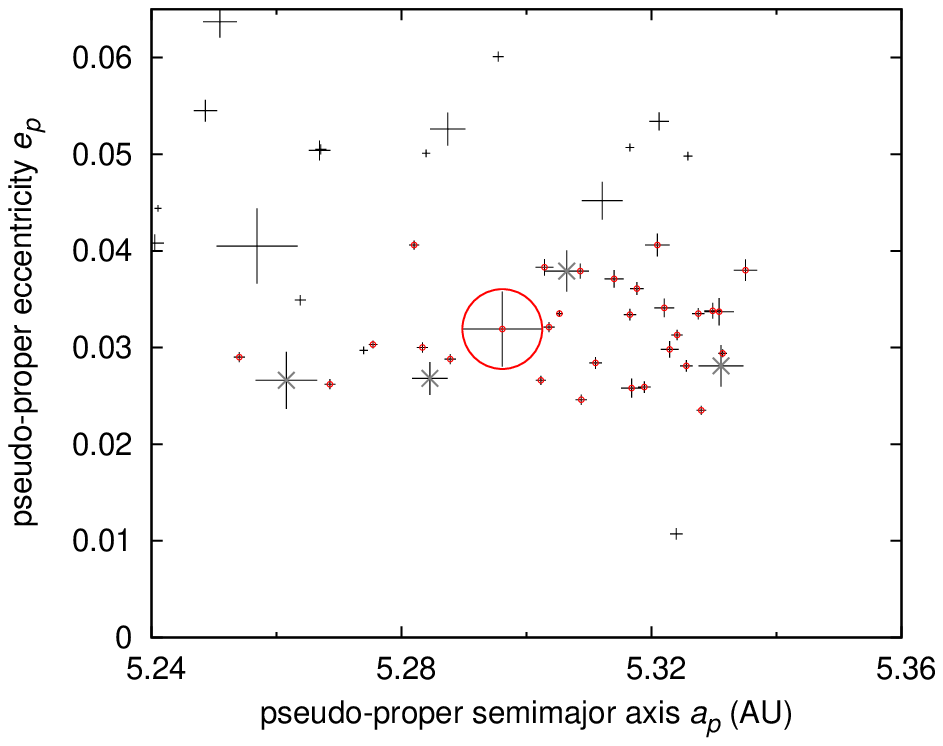}
\includegraphics[width=6cm]{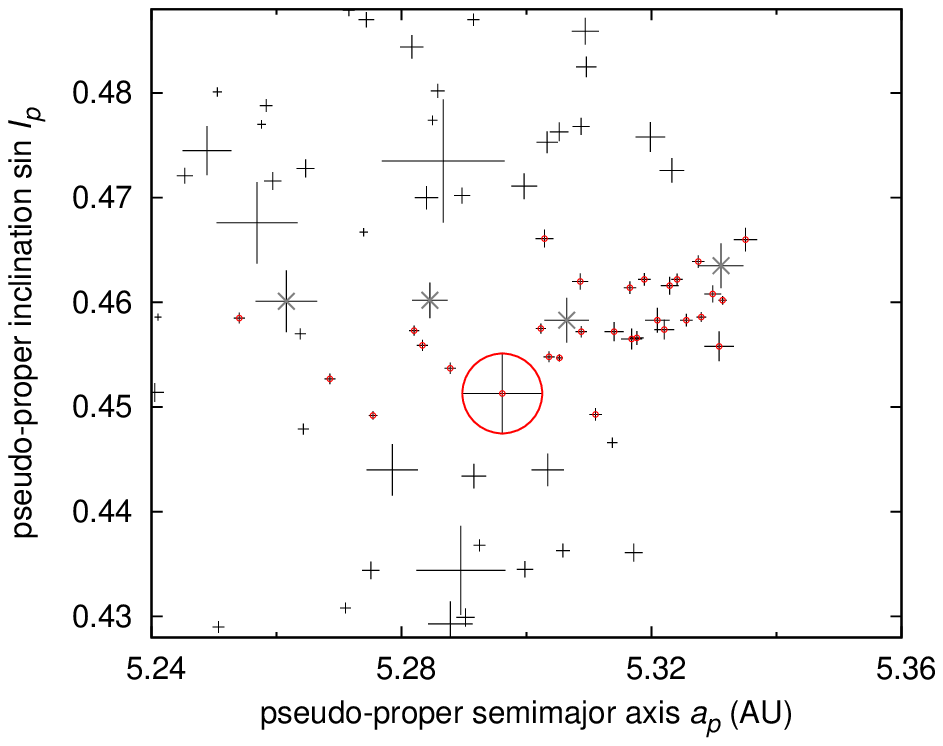}
\caption{A detail of the $L_5$ Trojan population where the Ennomos group is visible.
Left panel: resonant semimajor axis~$a$ vs eccentricity~$e$.
Only asteroids occupying the same range of inclinations as the Ennomos group
$\sin I \in (0.448, 0.468)$ are plotted to facilitate a comparison with
the density of surroundings space (background).
The sizes of plus signs are proportional to diameters of the asteroids.
Probable family members are denoted by small red circles
and probable interlopers by small grey crosses.
Right panel: $a$ vs inclination $\sin I$,
with range of eccentricities $e \in (0.02, 0.045)$.}
\label{trojans_L5_arer_sizes_Ennomos}
\end{figure*}

The cluster around (4709) Ennomos can be recognised for a wide
interval of cut--off velocities $v_{\rm cutoff} \in (69, 129)\,{\rm m}/{\rm s}$
when it stays compact and confined in inclinations $(\sin I = 0.451\hbox{ to }0.466)$.
Very probably, there are several interlopers, because we can count 4 to 10 asteroids
in the surroundings, i.e., in the same volume of the $(d, e, \sin I)$ space
(see Figure~\ref{trojans_L5_arer_sizes_Ennomos}). Since small bodies dominate
the Ennomos group we suspect large bodies might be actually interlopers.

A very intriguing feature is a high albedo of (4709) Ennomos $p_V \simeq 0.15$
measured by Fern\'andez et al. (2003). Apart from other explanations, the authors
speculated it may result from a recent impact which covered the surface with pristine ice.
If true the relation between the fresh surface and the collisional family
might be a unique opportunity to study cratering events.

We cannot exclude a possibility that (4709) Ennomos is actually
an interloper and the family is not related to it at all.
Nevertheless, our hypothesis is testable:
family members should exhibit a similarity in spectra and albedos.
The only information we have to date are SDSS colours for three
members: 98362, 2005~YG$_{204}$ are probably C-types
and 2005~AR$_{72}$ is a D-type.
In case new data become available we can remove interlopers
from our sample and improve our analysis.

The size distribution of the Ennomos group is barely constrained,
since small bodies are at the observational limit. Moreover,
removal of interlopers can change the SFD slope completely
(from $\gamma = -1.4$ to $-3.2$ or so).
The minimum parent body size is about $D_{\rm PB} \simeq 67\,{\rm km}$
if all members have high albedo $p_V = 0.15$.

%%%%%%%%%%%%%%%%%%%%%%%%%%%%%%%%%%%%%%%%%%%%%%%%%%%%%%%%%%%%%%%%%%%%%%

\subsection{Group denoted Aneas}

The Aneas group looks like a middle portion of the $L_5$ cloud
with approximately background density. It spans whole range
of semimajor axes, as background asteroids do.

The minimum size of a hypothetical parent body is $D_{\rm PB} = 160\hbox{ to }170\,{\rm km}$
(for albedo $p_V = 0.055\hbox{ to }0.041$). This size is very large
and an impact disruption of such body is less probable (see Section~\ref{sec:collisions}).
The size-frequency distribution is shallow, with approximately
the same slope as background.

According to Roig et al. (2008) the colours are rather homogeneous
and correspond to D-types, with $\simeq10\,\%$ of probable interlopers.

%%%%%%%%%%%%%%%%%%%%%%%%%%%%%%%%%%%%%%%%%%%%%%%%%%%%%%%%%%%%%%%%%%%%%%

\subsection{Group denoted 1988 RG$_{10}$}

The group around asteroid (11487) 1988 RG$_{10}$ again looks like
a lower portion of the $L_5$~cloud at low inclinations, with $\sin I \in (0.06, 0.1)$.
The SFD is steeper ($\gamma = -2.6\pm 0.1$) than surroundings in $L_5$
and the resulting parent body size $D \simeq 60\,{\rm km}$ is relatively small.
The colours seems heterogeneous (Roig et al. 2008) and we can
confirm this statement based on the new SDSS-MOC version~4 data.

%%%%%%%%%%%%%%%%%%%%%%%%%%%%%%%%%%%%%%%%%%%%%%%%%%%%%%%%%%%%%%%%%%%%%%

The remaining clusters
(Hektor, Teucer, Sinon, 1986~WD, Laertes, Asios, Polydoros, Panthoos, etc.)
may be characterised as follows:
  (i)~they have a density in $(d, e, \sin I)$ space comparable to that of background (surroundings);
 (ii)~when identified by the HCM their semimajor axes span the whole range of Trojan region;
(iii)~the slopes of their SFD's are comparable to the background;
 (iv)~they are often inhomogeneous with respect to colours (according to Roig et al. 2008).
These reasons lead us to a conclusion that these clusters
are not necessarily real collisional families.

%%%%%%%%%%%%%%%%%%%%%%%%%%%%%%%%%%%%%%%%%%%%%%%%%%%%%%%%%%%%%%%%%%%%%%
%%%%%%%%%%%%%%%%%%%%%%%%%%%%%%%%%%%%%%%%%%%%%%%%%%%%%%%%%%%%%%%%%%%%%%

\section{Long-term evolution of Trojan families}\label{sec:evolution}

\subsection{Evolution due to chaotic diffusion}

We try to model long-term evolution of the Eurybates family.
At first, we generate a synthetic family (consisting of 42 bodies)
by an impact disruption of the parent body with required size.
Then we integrate the synthetic family and compare it at particular
time to the observed Eurybates family. The time span of the integration is 4\,Gyr.

The main driving mechanism is slow {\em chaotic diffusion\/}
(the Yarkovsky effect is present but inefficient in the Trojan region).
Initially, the spread of inclinations of the synthetic family
is consistent with the observed one. On the other hand, the shape
in $(a, e)$ elements is clearly inconsistent.

\begin{figure*}
\centering
\includegraphics[width=5.8cm]{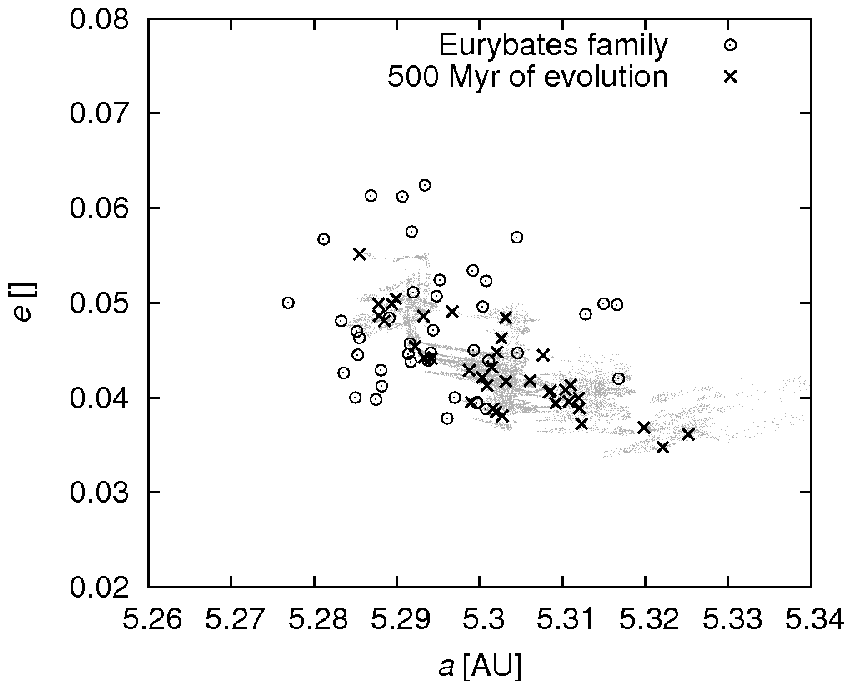}
\includegraphics[width=5.8cm]{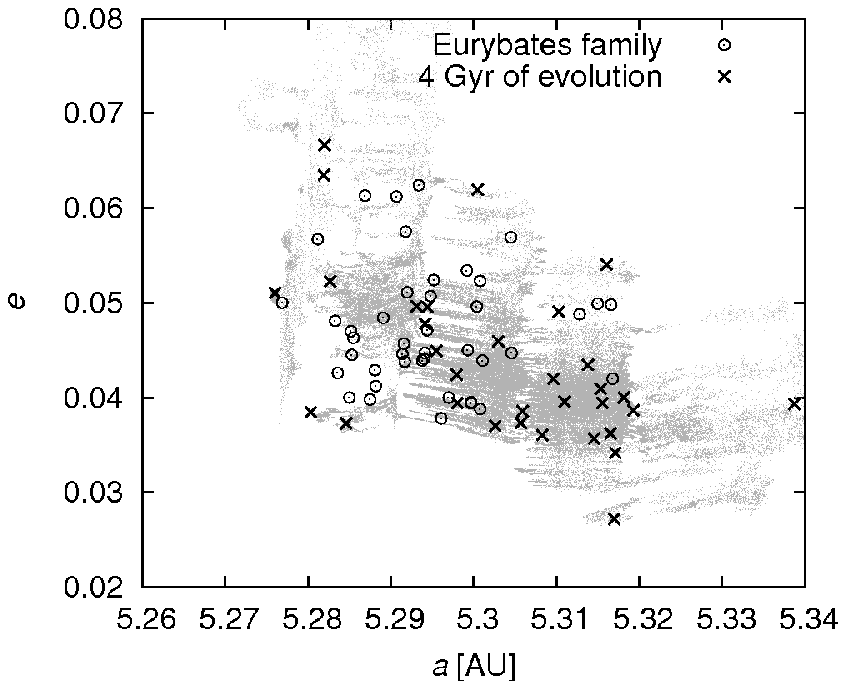}
\includegraphics[width=5.8cm]{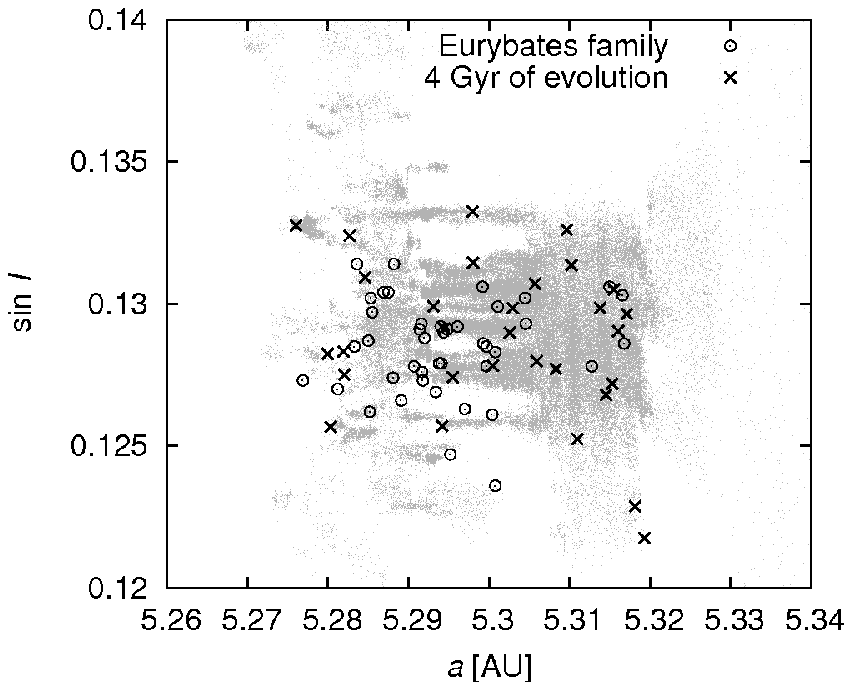}
\caption{Orbital evolution of the synthetic family and its comparison with the observed Eurybates family.
Left panel: the situation in the $(a, e)$ plane at 500\,Myr.
Middle panel: the situation after 4\,Gyr.
Chaotic diffusion disperses the synthetic family in course of time (see shaded tracks of particles).
Right panel: the $(a, \sin I)$ plane at the same time.
Inclinations evolve only barely.}
\label{eurybates-impye2_ae_500Myr}
\end{figure*}

Since the inclinations evolve only barely we focus on the evolution
of in the $(a, e)$ plane (see Figure~\ref{eurybates-impye2_ae_500Myr}).
The point is the synthetic family, namely the `filament' structure,
has to disperse sufficiently. After 500\,Myr it is still recognisable
but after 1\,Gyr of evolution it is not. So we may constrain
the age of the Eurybates family from 1~to 4\,Gyr.%
\footnote{We verified these estimates by a 2-dimensional Kolmogorov--Smirnov test of the $(a,e)$ distributions:
initially the KS distance is $D_{\rm KS} = 0.30$ and the probability $p_{\rm KS}({>}D) = 0.02$,
which means the distribution are incompatible.
At $t = 1\,{\rm Gyr}$, the values are $D_{\rm KS} = 0.20$ and $p_{\rm KS}({>}D) = 0.32$,
which indicates a reasonable match.}

% ^^^ these are for $e$ shifted by $+0.005$ and $a$ by $-0.005\,{\rm AU}$
% $D_{\rm KS} = 0.52$, $p_{\rm KS}(>D) = 3\cdot10^{-5}$,
% $D_{\rm KS} = 0.38$, $p_{\rm KS}(>D) = 3\cdot10^{-3}$.

A similar analysis for the Ennomos group indicates that
chaotic diffusion is faster in this region (given the large inclination)
and the most probable age thus seems to be from 1 to 2\,Gyr.
Beyond 2\,Gyr the inclinations of the synthetic family become
too large compared to the observed Ennomos group,
while the eccentricites are still compatible.

\begin{figure}
\centering
\includegraphics[width=6.5cm]{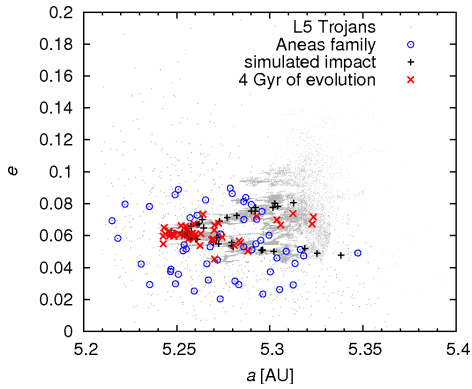}
\caption{Evolution of the synthetic family over 4\,Gyr versus the observed Aneas group.
Chaotic diffusion is slow and it seems impossible to match the large spread of the observed
group even after 4\,Gyr.}
\label{aneas-impye2_ae_4Gyr}
\end{figure}

We try to model Aneas and 1988 RG$_{10}$ groups too
(see Figure~\ref{aneas-impye2_ae_4Gyr}).
In these two cases, there is a strong disagreement between
our model and observations. The observed groups are much larger
and chaotic diffusion in respective regions is very slow.
Even after 4\,Gyr of orbital evolution, the synthetic family
remains too small.

The only free parameter which may substantially change our results
is the initial velocity distribution. Theoretically, the distribution
might have been strongly anisotropic.
However, we cannot choose initial velocities entirely freely,
since their magnitude should be comparable to the escape velocity
from the parent body, which is fixed by the size $D_{\rm PB}$
and only weakly dependent on a-priori unknown density $\rho_{\rm PB}$.

Another solution of this problem is possible if we assume families
are very old and they experienced perturbations due to planetary
migration.

%%%%%%%%%%%%%%%%%%%%%%%%%%%%%%%%%%%%%%%%%%%%%%%%%%%%%%%%%%%%%%%%%%%%%%

\subsection{Stability during planetary migration}\label{sec:migration}

The major perturbation acting on Trojans are {\em secondary resonances\/}
between the libration period~$P_{J1/1}$ of the asteroid in the J1/1 mean-motion resonance with Jupiter
and the period $P_{\rm 1J-2S}$ of the critical argument of Jupiter--Saturn 1:2 resonance (Morbidelli et al. 2005)
\begin{equation}
P_{\rm J1/1} = n P_{\rm 1J-2S}\,,
\end{equation}
where~$n$ is a small integer number.
Typical libration periods are $P_{\rm J1/1} \simeq 150\,{\rm yr}$
and $P_{\rm 1J-2S}$ changes as planets migrate
(it decreases because Jupiter and Saturn recede from their mutual 1:2 resonance).%
\footnote{Another source of instability might be a secondary resonance
with $P_{\rm 2J-5S}$ (the so called Great Inequality period)
thought it is weaker than $P_{\rm 1J-2S}$.
We find no asteroids perturbed by secondary resonances connected
with $P_{\rm 3J-7S}$ or $P_{\rm 4J-9S}$ which are present `en route'.
Neither Uranus nor Neptune play an important role.}

\begin{figure}
\centering
\includegraphics[width=8.3cm]{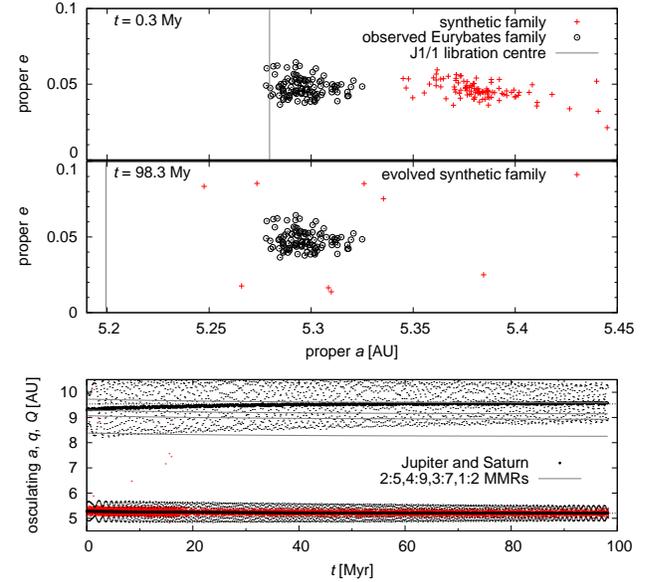}
\caption{Evolution of a synthetic family during late phases of planetary migration ($\tau_{\rm mig} = 30\,{\rm Myr}$ in this case).
Top panel: the state at 0\,Myr, middle: 100\,Myr, bottom: the respective orbital evolution of Jupiter and Saturn.
The family is almost destroyed and it is definitely incompatible
with the observed Eurybates family.}
\label{eurybates-mig3_DAMPE_ae_at_00000300}
\end{figure}

All synthetic families are strongly unstable when $P_{\rm 1J-2S} \simeq 150\,{\rm yr}$
and even during later stages of migration with $P_{\rm 1J-2S} \simeq 75\,{\rm yr}$
the eccentricities of family members are perturbed too much
to match the observed families like Eurybates or Ennomos (see Figure~\ref{eurybates-mig3_DAMPE_ae_at_00000300}).
There are practically no plausible migration scenarios -- regardless of time scale $\tau_{\rm mig}$ --
which would produce a sufficiently compact group, unless Jupiter and Saturn are almost
on their current orbits. We tested $\tau_{\rm mig} = 0.3, 3, 30\,{\rm Myr}$
and even for $\Delta a_{\rm J} \equiv a_{\rm Jf} - a_{\rm Ji}$ as small as $-0.08\,{\rm AU}$
and $\Delta a_{\rm S} = +0.25\,{\rm AU}$ the perturbation was too strong.
The reason is that one has to avoid $n=2$ secondary resonance to preserve
a low spread of a synthetic family.

Let us conclude if any of Trojan families was created during planetary migration
and if the migration was smooth (exponential) the family cannot be visible today.
However, we cannot exclude a possibility that final stages
of migration were entirely different, e.g., similar to the `jumping-Jupiter' scenario (Morbidelli et al. 2010).

% libration periods:
% J2/1 (Zhonnguos) - 440 yr
% J3/2 (Hildas)    - 270 yr
% J1/1 (Trojans)   - 150 yr

% tesne po 1:2 (a_J = 5.28 -> 5.20 AU, a_S = 8.62 -> 9.55 AU): ... (nepocital jsem)

% stredni faze migrace (r32-impmig2: a_J = 5.28 -> 5.20 AU, a_S = 8.82 -> 9.55 AU):

% 1J-2S - 150 yr az  60 yr   <- na zacatku jsou proto nestabilni Trojani! <- sekundarni rezonance 1:1 <- v Moridelli et al. (2005) to je spatne?!
% 2J-5S -  20 yr az 

% pozdni faze migrace (eurybates-mig3: a_J = 5.28 -> 5.20 AU, a_S = 9.30 -> 9.55 AU):

% 1J-2S                    - 80 az    60 yr <- v uvahu prichazi sekundarni rezonance 1:2, ALE telesa ubyvaji az chvili POTOM
% 2J-5S (Great Inequality) - 90 az ~2000 yr <- zde probehla s. r. 1:1 kraticce PO ubytku teles
%                                           <- aktualne 880 yr! (presvihnul jsem)

% skoro BEZ migrace: viz eurybates-mig4f (a_J = 5.22 -> 5.20, a_S = 9.525 -> 9.535 AU).

%%%%%%%%%%%%%%%%%%%%%%%%%%%%%%%%%%%%%%%%%%%%%%%%%%%%%%%%%%%%%%%%%%%%%%

\subsection{Families lost by the ejection of fragment outside the resonance}

\begin{figure}
\centering
\includegraphics[width=6.0cm]{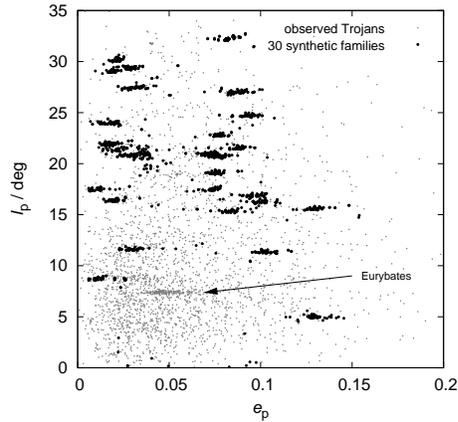}
\caption{Proper eccentricities and inclinations of 30~{\em synthetic\/} families (black dots),
which originated near the border of stable libration zone,
compared to the observed L4 Trojans (gray dots).}
\label{ballistic_transport_ei}
\end{figure}

We studied a possibility that some families cannot be identified because the
breakup occurred at the outskirts of the stable libration zone
and some fragments were ejected outside the J1/1 resonance.
We thus chose 30~largest asteroids near the edge of the $L_4$~libration zone
and we simulated breakups of these asteroids which create families with 30 fragments each.
We assumed the diameter of all parent bodies equal to $D_{\rm PB} = 100\,{\rm km}$
and their density $\rho_{\rm PB} = 1.3\,{\rm g}\,{\rm cm}^{-3}$.
The breakups always occurred at the same geometry $f_{\rm imp}=0^\circ$, $\omega_{\rm imp} = 30^\circ$. 
After the breakup, we calculated proper elements of the family members
and plotted their distribution (see Figure~\ref{ballistic_transport_ei}).
We can see all 30 synthetic families can be easily identified.
In most cases, more than 95\,\% of family members remained within the stable libration zone.
We can thus conclude that the ejection of fragments outside the libration zone
does not affect the number of observed families among Trojans.

%%%%%%%%%%%%%%%%%%%%%%%%%%%%%%%%%%%%%%%%%%%%%%%%%%%%%%%%%%%%%%%%%%%%%%

\subsection{Collisional rates}\label{sec:collisions}

We can estimate collisional activity by means of a simple stationary model.
Trojan--Trojan collisions play a major role here, because Trojans are detached
from the Main Belt.
In case of Eurybates, the target (parent body) diameter $D_{\rm target} = 110\,{\rm km}$,
the mean impact velocity $V_{\rm imp} = 4.7\,{\rm km}/{\rm s}$ (Dell'Oro et al. 1998),
the strength $Q^\star_D = 10^5\,{\rm J}/{\rm kg}$ (Benz \& Asphaug 1999)
and thus the necessary impactor size (Bottke et al. 2005)
\begin{equation}
d_{\rm disrupt} = \left({2 Q^\star_D / V_{\rm imp}^2}\right)^{1/3} D_{\rm target}
\simeq 23\,{\rm km}\,.\label{eq:d_disrupt}
\end{equation}
Number of ${\ge}23\,{\rm km}$ projectiles among $L_4$ Trojans is $n_{\rm project} = 371$
and we have $n_{\rm target} = 8$ available targets.
An intrinsic collision probability for Trojan--Trojan collisions
$P_{\rm i} = 7.8\times 10^{-18}\,{\rm km}^{-2}\,{\rm yr}^{-1}$ (Dell'Oro et al. 1998)
and corresponding frequency of disruptions is
\begin{equation}
f_{\rm disrupt} = P_{\rm i} {D_{\rm target}^2\over 4} n_{\rm project} n_{\rm target}
\simeq 7\cdot 10^{-11}\,{\rm yr}^{-1}\,.\label{eq:f_disrupt}
\end{equation}
Over the age of the Solar System $T_{\rm SS} \simeq 4\,{\rm Gyr}$ (after the LHB),
we have a very low number of such events
$n_{\rm events} = T_{\rm SS} f_{\rm disrupt} \simeq 0.28$.
This number seems to be in concert with only one $D \ge 100\,{\rm km}$ family
currently observed among Trojans.%
\footnote{A similar stationary estimate valid for the Main Asteroid Belt
gives the number of events~$12$ while the number of observed families
with $D_{\rm PB} \gtrsim 100\,{\rm km}$ is about~20 (Durda et al. 2007).
These two numbers are comparable at least to order-of-magnitude.}
In a less-likely case, the material of the Eurybates parent body was very weak
and its strength may be at most one order of magnitude lower, $Q^\star_D \simeq 10^4\,{\rm J}/{\rm kg}$
(see Leinhardt \& Stewart 2009, Bottke et al. 2010).
We then obtain $d_{\rm disrupt} \simeq 10\,{\rm km}$ and $n_{\rm events} \simeq 1.0$,
so the conclusion about the low number of expected Trojan families remains essentially the same.

The parent body of Aneas group is 1.5 larger and consequently the resulting
number of events is more than one order of magnitude lower.
On the other hand, clusters with smaller parent bodies ($D_{\rm PB} \ll 100\,{\rm km}$)
or significantly weaker ($Q^\star_D \ll 10^{5}\,{\rm J}/{\rm kg}$)
might be more frequent.

During the Late Heavy Bombardment epoch we may assume a substantial
increase of collisional activity (Levison et al. 2009). Hypothetical
old families were however probably `erased' due to the late phases
of planetary migration (see Section~\ref{sec:migration})
unless the migration time scale for Jupiter and Saturn
was significantly shorter than the time scale of the impactor flux
from transneptunian region which is mainly controlled by the
migration of Uranus and Neptune.

%% see src/disruption_Eurybates.plt and .out

%%%%%%%%%%%%%%%%%%%%%%%%%%%%%%%%%%%%%%%%%%%%%%%%%%%%%%%%%%%%%%%%%%%%%%
%%%%%%%%%%%%%%%%%%%%%%%%%%%%%%%%%%%%%%%%%%%%%%%%%%%%%%%%%%%%%%%%%%%%%%

\section{Conclusions}\label{sec:conclusions}

Increasing number of Trojan asteroids with available proper elements
enables us to get new insights into this important population.
Essentially, new faint/small asteroids filled the `gaps'
in the proper-element space between previously known clusters
and today it seems most clusters are rather comparable to background.
One should be aware that the number of families among Trojans
may be low and one should not take the number of $\simeq 10$ families
as a rule.

Only the C-type Eurybates family fulfils all criteria to be considered
a collisional family. This is probably also true for the newly discovered
Ennomos group. Moreover, there might be a potentially interesting
relation between the high-albedo surface of (4709) Ennomos and the
collisional family thought we do not have enough data yet to prove
it independently (by colours, spectra or albedos).

Note there may exist clusters among Trojans which are not
of collisional origin. They may be caused by:
(i)~differences in chaotic diffusion rates;
(ii)~$a/e/I$-dependent efficiency of original capture mechanism; or
(iii)~it may somehow reflect orbital distribution in source regions.

We cannot exclude a hypothetical existence of old families
which were totally dispersed by dynamical processes,
e.g., by perturbations due to planetary migration which
is especially efficient in the Trojan region.

Finally, note there seem to be no D-type families anywhere
in the Solar System --- neither in the Trojan region,
nor in the J3/2 (Bro\v z et al. 2011) and Cybele regions (Vokrouhlick\'y et al. 2010).
May be, the D-type parent bodies are too weak and the target
is completely pulverized during a collision?
This might have important implications for collisional models
of icy bodies.

%%%%%%%%%%%%%%%%%%%%%%%%%%%%%%%%%%%%%%%%%%%%%%%%%%%%%%%%%%%%%%%%%%%%%%

\section*{Acknowledgements}
We thank David Vokrouhlick\'y, David Nesvorn\'y, Alessandro Morbidelli and William F. Bottke
for valuable discussions on the subject.
We also thank Lenka Trojanov\'a for the pricipal component analysis of the SDSS-MOC4 data
and Gonzalo Carlos de El\'ia for his review which helped to improve the final version of the paper.
The work has been supported by the Grant Agency of the Czech Republic (grant 205/08/P196)
and the Research Program MSM0021620860 of the Czech Ministry of Education.
We acknowledge the usage of computers of the Observatory and Planetarium in Hradec Kr\'alov\'e
and Observatory and Planetarium of Prague.

%%%%%%%%%%%%%%%%%%%%%%%%%%%%%%%%%%%%%%%%%%%%%%%%%%%%%%%%%%%%%%%%%%%%%%

%%%%%%%%%%%%%%%%%%%%%%%%%%%%%%%%%%%%%%%%%%%%%%%%%%%%%%%%%%%%%%%%%%%%%%%%

\bsp

\label{lastpage}

\end{document}